\documentclass[a4paper,showpacs,twocolumn,amsmath,amssymb,floatfix,prl,preprintnumbers,footinbib]{revtex4}
\usepackage{graphicx}
\usepackage{dcolumn}
\usepackage{bm}

\begin{document}


\title{Top-gate defined double quantum dots in InAs nanowires}
\author{A.~Pfund, I.~Shorubalko, R.~Leturcq and K.~Ensslin}

\address{Solid State Physics Laboratory, ETH Z\"urich, 8093 Z\"urich, Switzerland\\
E-mail: leturcq@phys.ethz.ch }

\begin{abstract}
We present low temperature transport measurements on double quantum dots in InAs nanowires grown by metal-organic vapor phase epitaxy. Two dots in series are created by lithographically defined top-gates with a procedure involving no extra insulating layer. We demonstrate the full tunability from strong to weak coupling between the dots. The quantum mechanical nature of the coupling leads to the formation of a molecular state extending over both dots. The excitation spectra of the individual dots are observable by their signatures in the nonlinear transport.
\end{abstract}

\maketitle

Semiconductor nanowires (NWs) are gaining interest as possible building blocks in new bottom-up nanoelectronics and as versatile systems for transport studies in reduced dimensions. Self-assembled or catalyzed epitaxial growth has been realized with large control over structural parameters such as diameter, length, crystal structure and position of the nanowires \cite{XiaY01,Samuelson04}. Among the III-V semiconductors, InAs is characterized by a small effective mass, a large effective $g^*$-factor and strong spin-orbit interaction. InAs is therefore considered as a promising material for the investigation of quantum mechanical and spin related phenomena in electronic transport. Single charges and spins can be studied in quantum dots \cite{Elzerman03,Fujisawa02,Ono02} and are envisioned as candidates for quantum bits (qubits) in solid-state based quantum computers \cite{Loss01}. Coupled systems like double quantum dots appear to be especially promising in this context due to the possibility to control the entanglement of spins by tuning the exchange coupling between the dots \cite{DiVincenzo00,Burkhard99}. Reliable readout and manipulation schemes for qubit states in dots have been demonstrated  \cite{Petta04,Johnson05}. A crucial prerequisite for these schemes is a good tunability of the tunnel barriers, energy levels and particularly the quantum mechanical coupling of the two dots.

\begin{figure}
\centering
\includegraphics[angle=270,width=.45\textwidth]{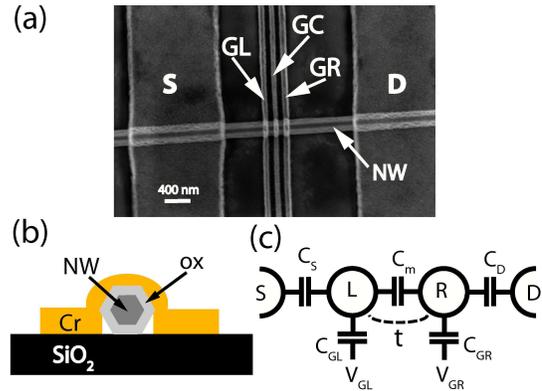}
\caption{(Color online) (a) Scanning electron micrograph of a device similar to the one measured. Source (S) and drain (D) ohmic contacts to the nanowire (NW) are defined by optical lithography. Top-gates GL, GR and GC have width and spacing of about 70 nm. (b) Schematic showing the NW with hexagonal cross section. The top-gates are isolated by a native surface oxide layer (ox). (c) Model for the analysis of the double dot system. $C_{GL},C_{GR}$ are the capacitances coupling the gate voltages to the electrochemical potential in the left (L) and right (R) dot respectively. $C_S,C_D$ represent the electrostatic action of source and drain potential. $C_m$ is the mutual capacitive coupling between the two dots. A tunnel coupling $t$ between the two dots is also taken into account.}
\end{figure}

Here we describe a technologically simple and reliable process to create top-gate defined double quantum dots in InAs NWs. We present transport measurements demonstrating the full tunability of the interdot coupling, ranging from an artificial ``quantum-dot molecule'' to weakly capacitively coupled quantum dots. Beyond that, nonlinear transport measurements at finite bias clearly reveal the excitation spectrum of the individual dots.

Following the pioneering experiments published in Refs.~\onlinecite{Hiruma01} and \onlinecite{Seifert04}, we grow InAs NWs on $\langle 111 \rangle$B oriented GaAs substrates by metal-organic vapor phase epitaxy. The growth process is catalyzed by monodisperse Au particles with diameters of 40 nm. The resulting NWs are between 5 and 10 $\mu\textrm{m}$ long and have diameters around 100 nm. They emerge with crystalline facets and hexagonal cross-section. After growth, the NWs are transferred to a highly doped Si substrate covered by 300 nm of $\textrm{SiO}_2$. No predefined markers are needed, since the NWs can be located optically and contacts are defined by a single optical lithography step. Before metalization with subsequent layers of Ti (20 nm) and Au (180 nm), the contact areas are simultaneously etched and passivated with a diluted solution of (NH$_4$)$_2$S$_\textrm{x}$ \cite{Oigawa01}. The resulting contacts are of good quality with contact resistances below 100 $\Omega$. Finally, top-gate fingers are created by electron beam lithography, followed by deposition of 6 nm Cr, 66 nm Au and standard lift-off. Note that, for this step, no etching and passivation is done, keeping a native surface oxide layer of 1--3 nm thickness formed after our growth process \cite{Chimia06}. We find that this oxide is suitable to electrically isolate the top-gates for voltages in the range of at least $\pm 1$ V. As a result, ``half wrap-around-gates'' are created with no additional deposition of an insulating layer between the NW and the top-gates \cite{vanDam01}. Details of the fabrication process can be found elsewhere \cite{Chimia06}. This approach is technologically simple compared to previously demonstrated methods, where the NW is locally gated with predefined back-gate fingers \cite{Fasth01}. A scanning electron microscope image of the device and a schematic of the gates is shown in Fig.$\,$1~(a) and (b).

We performed electrical measurements on two independent devices showing comparable behavior. Data of one sample measured in a dilution refrigerator at a base temperature of 30 mK are presented. Ungated pieces of the NW show a resistance of around $10\,\textrm{k}\Omega$ per $\mu\textrm{m}$ length. With all top-gates put to 0 V, the resistance between source S and drain D is above $1\,\textrm{M}\Omega$. Applying +0.5 V to all top-gates compensates for the presence of the gate fingers and recovers the resistance of the ungated device. Two quantum dots in series can be formed by reducing the voltages on gates GL, GC and GR. Fig.$\,$2 shows plots of the source-drain current $I_{SD}$ at $V_{SD}=140$ $\mu\textrm{V}$ as a function of the voltages on left ($V_{GL}$) and right ($V_{GR}$) top-gates for 3 different values of the center gate voltage ($V_{GC}$). In Fig.$\,$2(a) no significant barrier is induced by GC and a single quantum dot is formed between GL and GR. The current exhibits Coulomb oscillations: peaks occur when the electrochemical potential for the transition of an N electron state to the N+1 electron state is aligned with the electrochemical potentials $\mu_S$ and $\mu_D$ in the contacts. Reducing $V_{GC}$ leads to the formation of a double-well potential and induces a redistribution of the N electrons between the two valleys (Fig.$\,$2(b)). The electron numbers in the left and right dots are denoted as $n$ and $m$ respectively. For even smaller $V_{GC}$, we obtain the characteristic honeycomb pattern for the charge stability diagram of the double dot \cite{vdWiel01}. Within each hexagon of Fig.$\,$2(c), the charge configuration $(n,m)$ of the double dot is fixed. Elastic electron transport leading to high current is only possible at the corners of the hexagons marking the so-called triple points. Here, the electrochemical potentials of the $n\rightarrow n+1$ electron transition in the left dot and the $m\rightarrow m+1$ electron transition in the right dot are aligned with  $\mu_S$ and  $\mu_D$. The series shown in Fig.$\,$2 demonstrates the tunability from one large single dot to two mainly capacitively coupled dots.

\begin{figure}
\centering
\includegraphics[width=.45\textwidth]{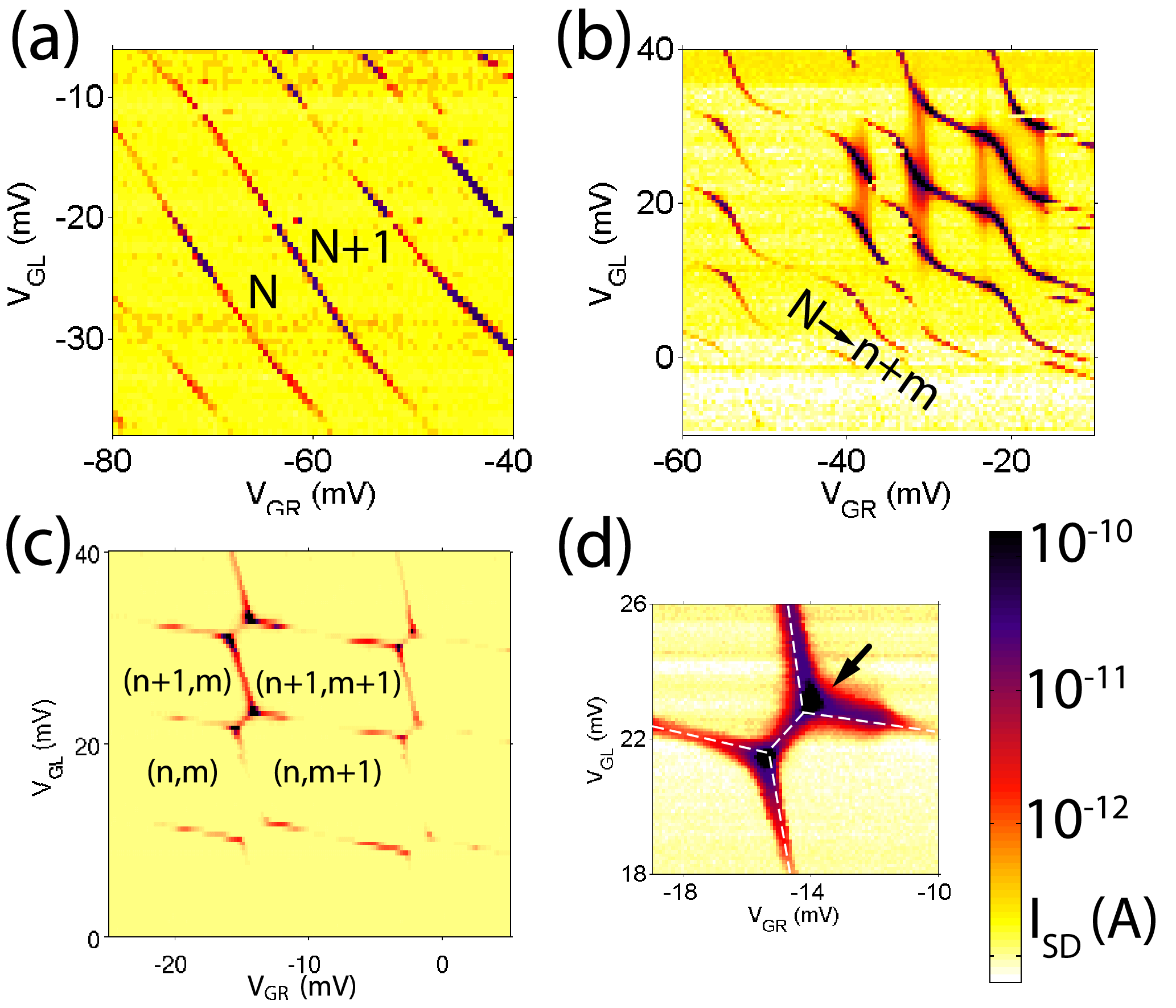}
\caption{(Color online) Source-drain current $I_{SD}$ as a function of left and right gate voltages for different center gate voltages at fixed bias $V_{SD}=140\,\mu\textrm{V}$. (a) At $V_{GC}=+200$ mV, an effective single dot is formed. The total number of electrons between the Coulomb peaks is fixed ($N, N+1, \ldots$). (b) $V_{GC}=0$ mV. Electrons are redistributed in the double well potential. (c) $V_{GC}=-120$ mV yields the charge-stability diagram for two capacitively coupled dots. (d) Zoom around one pair of degeneracy points. The deviation from the capacitive model (dashed lines) is clearly visible (see arrow).}
\end{figure}

At finite bias, the degeneracy points between different charge configurations develop into triangular regions of increased conductance (Fig.$\,$3, \cite{comment_rearrangement}). Inside these triangles, current flows due to inelastic tunneling processes. With a model of purely capacitively coupled dots \cite{vdWiel01}, we can extract all capacitances (as defined in Fig.$\,$1(c)) of the double dot system. From the dimensions of the honeycomb cells ($\Delta V_{GL}, \Delta V_{GR}$), we obtain the capacitances of GL on the left dot (GR on the right dot) as $C_{GL}=|e|/\Delta V_{GL}=13.7\,\textrm{aF}$ ($C_{GR}=12.7\,\textrm{aF}$). Equating the sizes of the triangles ($\delta V_{GL}, \delta V_{GL}$) to the applied bias $V_{SD}$, the leverarms for conversion of gate-voltages into energies follow to be $\alpha_L=0.46$ and $\alpha_R=0.41$. The corresponding total capacitances of the left (right) dot, $C_{L}=C_{GL}/\alpha_L=29.7\,\textrm{aF}$ ($C_{R}=30.7\,\textrm{aF}$) imply single dot charging energies $E_C^L= e^2 C_R/ (C_L C_R - C_m^2) \approx e^2/C_L=5.4\,\textrm{meV}$ ($E_C^R\approx5.2\,\textrm{meV}$). These values are consistent with Coulomb blockade measurements on the individual dots between two neighboring gates 
(not shown here). Assuming spherical dots embedded in InAs ($\epsilon_r=15$), the capacitances imply radii $R=C_{L,R}/4\pi\epsilon_0\epsilon_r\approx 18\,\textrm{nm}$, which are consistent with the lithographic dimensions. Considering electron densities of around $10^{18}\,\textrm{cm}^{-3}$ in the NWs \cite{Chimia06}, we estimate the number of electrons to be less than 20 in the dots.
The mutual capacitance between the dots is related to the splittings of the triple points $\Delta V_{GL,R}^m$ by $\Delta V_{GL,R}^m=|e|C_m/C_LC_R$, which leads to $C_m\approx 2.8\,\textrm{aF}$.

Quantum mechanical tunnel coupling between the states in both dots induces the formation of molecular states which are extended over the whole double dot system \cite{Blick01}. The presence of tunneling is clearly indicated by the rounded corners of the honeycombs in the vicinity of the triple points (Fig.$\,$2(c,d)). A Heitler-London picture for two degenerate quantum levels has proven to give a good qualitative and quantitative description for the curvature around the degeneracy points \cite{Ziegler00,Hatano05}. In this model, the spacing of two neighboring triple points is approximately given by $\gamma_m+2t$, where $t$ is the tunnel coupling between the levels and $\gamma_m=2e^2C_m/(C_L C_R - C_m^2)$ is the total capacitive splitting of the triple points \cite{Ziegler00}. Neglecting the tunnel coupling therefore leads to an overestimation of the capacitive coupling energy. A zoom around one pair of triple points is shown in Fig.$\,$2(d). The deviation from the purely capacitive model (dashed lines) is clearly visible. From the magnitude of the anticrossing, we estimate a tunnel coupling of $t=0.27\,\textrm{meV}$. The extracted capacitive interdot coupling $\gamma_m=0.87\,\textrm{meV}$ implies $C_m\approx2.5\,\textrm{aF}$. This is consistent with the capacitive analysis described above, considering the expected overestimation in the latter case. The strong tunnel coupling can also explain the shape of the upper triangles in Fig.$\,$3(a): a large current can flow out of the triangles defined by the bias voltage, due to sequential tunneling through superposed double dot states of the form $|E\rangle = c_1|n+1,m\rangle+c_2|n,m+1\rangle$ with complex coefficients $c_1,c_2$ \cite{Graeber2006}.

Another impact of quantum mechanics on transport through the double dot appears in the fine structure of the finite bias triangles. As can bee seen in Fig.$\,$3(b), stripes of high current emerge on the background of inelastic transport. These lines arise due to elastic transport when levels originating from excited dot states are in resonance with $\mu_S$ \cite{vdWiel01}. Note that in this case, the features are explained with excited orbital states in the two individual dots rather than by molecular states. These states give rise to different electrochemical potentials for a transition  $(n,m)\rightarrow (n+1,m)$. A possible scenario is described by the scheme in Fig.$\,$3(c).  At the lower left corner (A) of the triangles, the ground state levels are aligned with each other and with $\mu_S$. Tuning the top-gates along the line from point A to B, the potential in the right dot is lowered until the ground state is in resonance with $\mu_D$. Varying the gates along the connection from A to C pulls down the levels in both dots simultaneously. At the points labeled with $X$ (and similar for $X^\prime$), chemical potentials for transitions involving excited states (dashed lines in Fig.$\,$3(c)) enter the bias window giving rise to an increased current.

\begin{figure}
\centering
\includegraphics[width=.45\textwidth]{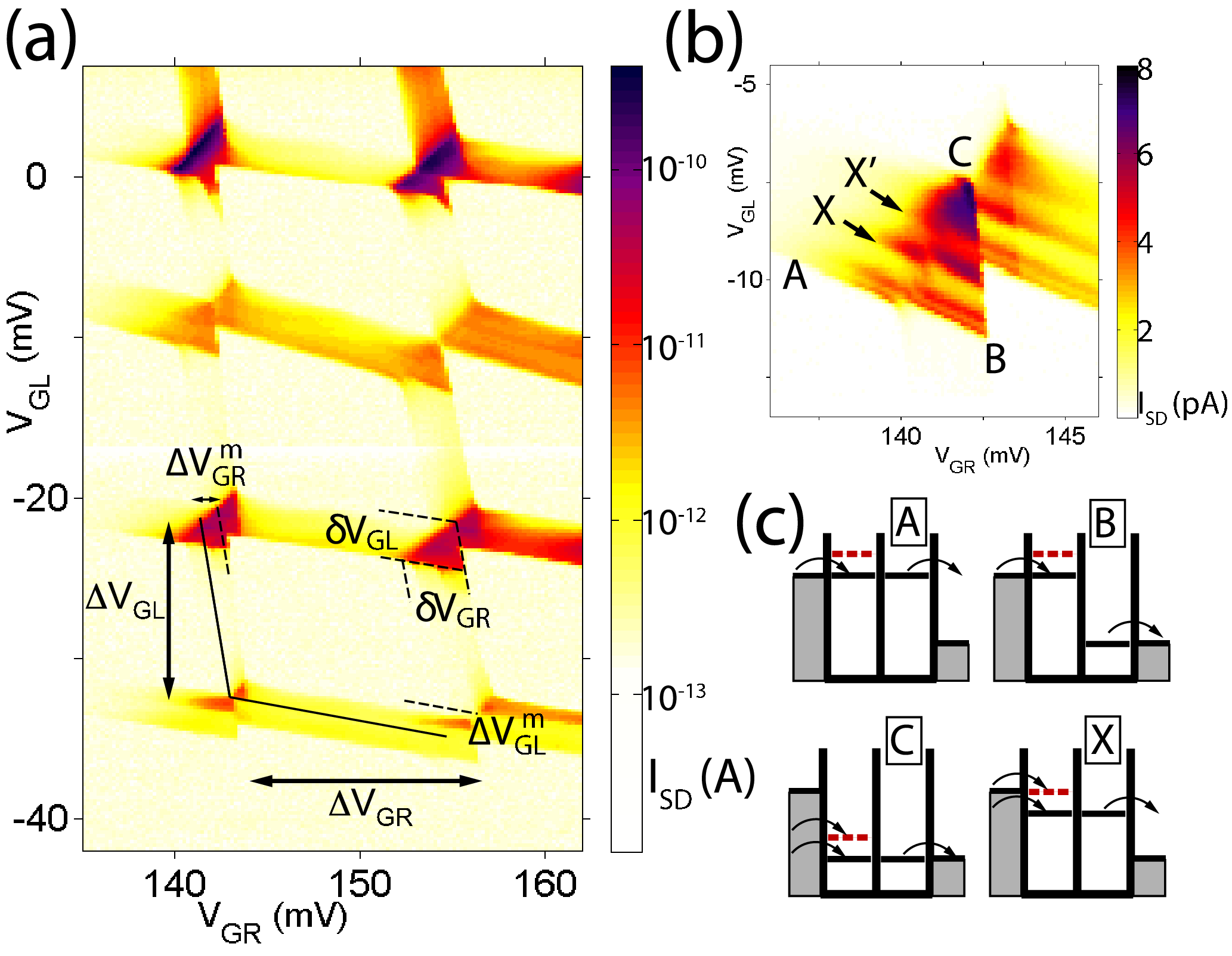}
\caption{(Color online) (a) Measurement of $I_{SD}$ as in Fig.$\,$2, but at finite bias $V_{SD}=1.2$mV. $V_{GC}=-120$ mV.
From the dimensions of the honeycombs and bias-triangles, all capacitances of the system can be extracted (see text). (b) Zoom into one pair of degeneracy points. Lines starting at points $X,X^\prime$ are explained by excited states. (c) Possible scenario for the level alignment at the points A, B, C, X.}
\end{figure}

In conclusion, we have realized top-gate defined double quantum dots in InAs nanowires. The full tunability of the interdot coupling could be demonstrated. We analyzed all characteristic capacitances of the system in a classical electrostatic model and quantified the tunnel coupling leading to the formation of a molecular state in the double dot. The presence of excited quantum states is observed in the transport at finite bias. This system therefore appears to be suitable for further experiments on controlling entanglement by tuning the exchange coupling between single spins.

We thank M.~T. Borgstr\"om and E. Gini for advises in the growth. We acknowledge financial support from the ETH Zurich, and IS thanks the European Commission for a Marie-Curie fellowship.


\begin{thebibliography}{10}
\bibitem{comment_rearrangement}
In this measurement, a charge rearrangement affecting only GR shifted the necessary gate voltage on the right gate by approximately 150 mV compared to Fig.$\,$2. The center gate, which determines the interdot coupling, has not been changed.
\end{thebibliography}
\end{document}